%Paper: hep-th/9212150
%From: A.Kirillov@newton.cam.ac.uk
%Date: Thu, 24 Dec 92 13:44 GMT

%-----------------------------------------------------------------
%-----------------------------------------------------------------
%
% Anatol N.Kirillov
%
% Dilogarithm identities, partitions and spectra in conformal field
% theory, part I.
%
% December 1992
%-------------------------------------------------------------------
%------------------------------------------------------------------
%
% Fill in the data (title, authors name, etc.) of your manuscript.
% The places to be filled in occur immediately after the symbols '!',
% hence you can search for '!' one by one.
\font\tenbf=cmbx10

\font\eightrm=cmr8
\font\eightit=cmti8

\def\sectiontitle#1\par{\vskip0pt plus.1\vsize\penalty-250
 \vskip0pt plus-.1\vsize\bigskip\vskip\parskip
 \message{#1}\leftline{\tenbf#1}\nobreak\vglue 5pt}
\def\qed{\hbox{${\vcenter{\vbox{
    \hrule height 0.4pt\hbox{\vrule width 0.4pt height 6pt
    \kern5pt\vrule width 0.4pt}\hrule height 0.4pt}}}$}}
%\TagsOnRight
\hsize=5.0truein
\vsize=7.8truein
\parindent=15pt
\nopagenumbers
\baselineskip=10pt
\line{\eightrm
Preliminary version
\hfil}
\vglue 5pc
\baselineskip=13pt
\headline{\ifnum\pageno=1\hfil\else%
{\ifodd\pageno\rightheadline \else \leftheadline\fi}\fi}
\def\rightheadline{\hfil\eightit
Dilogarithm identities, partitions and spectra in conformal field theory
%! Running title (odd page)
\quad\eightrm\folio}
\def\leftheadline{\eightrm\folio\quad
\eightit
Anatol N. Kirillov
%! Running author (even)
\hfil}
\voffset=2\baselineskip
\centerline{\tenbf
DILOGARITHM \hskip 0.1cm IDENTITIES, \hskip 0.1cm PARTITIONS \hskip 0.1cm AND
 }
\centerline{\tenbf
SPECTRA \hskip 0.1cm IN \hskip 0.1cm CONFORMAL \hskip 0.1 cm
FIELD \hskip 0.1cm THEORY,\hskip 0.1cm I }
\vglue 24pt
\centerline{\eightrm
ANATOL N. KIRILLOV%
\footnote*{This work is supported by SERC Grant GR59981}
%! AUTHOR:
%\footnote"$^*$"{\eightrm \baselineskip=10pt
%! present address
% }
}
\baselineskip=12pt
\centerline{\eightit
Isaac Newton Institute for Mathematical Sciences,
%! Use the permanent address (University Name, etc.)
}
\baselineskip=10pt
\centerline{\eightit
20 Clarkson Road, Cambridge, CB3 OEH, U.K.
}
\baselineskip=12pt
\centerline{\eightit
and }
\baselineskip=12pt
\centerline{\eightit
Steklov Mathematical Institute,
%! Use the permanent address (University Name, etc.)
}
\baselineskip=10pt
\centerline{\eightit
Fontanka 27, St.Petersburg, 191011, Russia
%! The permanent address (City, State ZIP/Zone, Country)
}
%More than two authors, see the sample prints.
\vglue 20pt
\centerline{  }
%\centerline{\eightrm Received \quad October 15, 1991}
%\vglue 16pt
%\baselineskip=10pt
%\centerline{\eightrm Revised  {\qquad\qquad} by Publisher)}
%\vglue 20pt
\centerline{\eightrm ABSTRACT}
{\rightskip=1.5pc
\leftskip=1.5pc
\eightrm\parindent=1pc
We prove new identities between the values of Rogers dilogarithm function and
describe a connection between these identities and spectra in conformal field
theory (part~I). We also describe the connection between asymptotical
behaviour of partitions of some class and the identities for Rogers
dilogarithm function (part II).
\vglue12pt}
\baselineskip=13pt
\overfullrule=0pt
%-------------------------------------------------------------------
%
%\input k1
%-------------------------------------------------------------------

\def\qed{\hfill$\vrule height 2.5mm width 2.5mm depth 0mm$}

{\bf Introduction.}
\bigbreak

The dilogarithm function $Li_2(x)$ defined for $0\le x\le 1$ by
$$Li_2(x):=\sum_{n=1}^{\infty }{x^n\over n^2}=-\int_0^x{\log (1-t)\over t}dt,
$$
is one of the lesser transcendental functions. Nonetheless, it has many
intriguing properties and has appeared in various branches of mathematics and
physics such as number theory (the study of asymptotic behaviour of partitions,
e.g. [RS], [AL]; the values of $\zeta$-functions in some special points [Za])
and algebraic $K$-theory (the Bloch group and a torsion in $K_3({\bf R})$ -
A.Beilinson, S.Bloch [Bl], A.Goncharov), geometry of hyperbolic
three-manifolds [Th], [NZ], [Mi], [Y], representation theory of Virasoro and
Kac-Moody algebras [Ka], [KW], [FS] and conformal field  theory (CFT).

In physics, the dilogarithm appears at first from a calculation of magnetic
susceptibility in the $XXZ$ model at small magnetic field [KR1], [KR2], [Ki1],
[BR]. More recently [Z], the dilogarithm identities (through the Thermodynamic
Bethe Ansatz (TBA)) appear in the context of investigation of $UV$ limit or the
critical behaviour of integrable 2-dimensional quantum field theories and
lattice models [Z], [DR], [K-M], [KBP], [Kl], [KP],...). Evenmore, it was
shown (e.g. [NRT]) using a method of Richmond and Szekeres [RS], that the
dilogarithm identities may be derived from an investigation of the asymptotic
behaviour of some characters of $2d$ CFT. Thus, it seems a very interesting
problem  to lift the dilogarithm identity in question to some identity between
the characters of certain conformal field theory. A partial solution of this
problem (without any proofs!) contained in [Te] and [KKMM].

One aim of this paper (part I) is to prove some new identities between the
values of Rogers dilogarithm function using the analytical methods (see [Le],
[Ki1]). Based on such identities we  give an answer on one question of W.Nahm
[Na]: how big may be the following abelian group
$${\cal W}:=\bigg\{\sum_i{n_iL(\alpha_i)\over L(1)}~|~n_i\in{\bf Z},~~
\alpha_i\in{\overline{\bf Q}}\cap{\bf R}~~{\rm for~all}~i\bigg\}\cap{\bf Q}~~?
$$
\medbreak

{\bf Theorem $\clubsuit$ .} The abelian group ${\cal W}$ coincides with
${\bf Q}$, i.e.
any rational number may be obtained as the value of some dilogarithm sum.

A proof of Theorem $\clubsuit$ follows from Proposition 4.5. We also give a
proof (and
the different generalizations) of an identity (3.1) from [NRT] (see our Theorem
3.1 and Proposition 5.4). Note the following ``reciprosity law'' for
dilogarithm sums (see Theorem 3.2)
$$s(j,n,r)+s(j,r,n)=nr-1.
$$
This is a consequence of the corresponding reciprosity law for the Dedekind
sums [Ra]. Note also that the author don't know any CFT interpretation for
the dilogarithm identities from Propositions 4.4 and 5.4.

Part II will be deal with the partition identities which''materialise'' the
dilogarithm ones.
%One aim of this paper is to prove some
%new identities between the values of Rogers dilogarithm function and to show
%that any rational number may be obtained as the value of some dilogarithm
%sum. More detailes and further results will appear in [Ki2].
\vskip 0.5cm

{\bf \S 1.~Definition~and~the~basic~properties~of~Rogers~dilogarithm.}

\bigbreak

Let us remind the definition of the Rogers dilogarithm function $L(x)$ for
\break
$x\in (0,1)$
$$
L(x)=-{1\over 2}\int_0^x\left[ {\log (1-x)\over x}+{\log x\over 1-x}\right]dx=
\sum_{n=1}^\infty {x^n\over n^2}+{1\over 2}\log x\cdot \log(1-x).
%~~0\le x\le 1.
\eqno (1.1)
$$

The following two classical results (see e.g. [Le2], [GM], [Ki1]) contain the
basic properties of the function $L(x)$.

\medbreak

{\bf Theorem~A}.~~The function $L(x)\in C^{\infty}((0,1))$ and satisfies the
following functional equations
$$\eqalignno{\hfil &1.~~L(x)+L(1-x)={\pi^2\over 6},~~0<x<1,
%L(1),~~
%x\in\bf R
& (1.2)\cr
&2.~~L(x)+L(y)=L(xy)+L\left({x(1-y)\over 1-xy}\right)+
L\left({y(1-x)\over 1-xy}\right),& (1.3)}
$$
where $0<x,~y<1$.

\medbreak

{\bf Theorem~B}.~~Let$f(x)$ be a function of class $C^3((0,1))$ and satisfies
the  relations (1.2) and (1.3). Then we have
$$ f(x)={\rm const}\cdot L(x)$$

We continue the function $L(x)$ on all real axis
${\bf R}={\bf R}^1\cup\{\pm\infty\}$ by the following rules
$$\eqalignno{
& L(x)={\pi^2\over 3}-L(x^{-1}),~~{\rm if}~~x>1, &(1.4)\cr
& L(x)=L\left({1\over 1-x}\right) -{\pi^2\over 6},~~{\rm if}~~x<0,& (1.5)\cr
& L(0)=0,~~L(1)={\pi^2\over 6},~~L(+\infty )={\pi^2\over 3},~~L(-\infty )=
-{\pi^2\over 6}.}
$$

The present work will concern with relations between the values of the
Rogers dilogarithm function at certain algebraic numbers. More exactly, let us
consider an abelian subgroup ${\cal W}$ in the field of rational numbers
${\bf Q}$
$$
{\cal W}=\left \{ \sum_i{n_iL(\alpha_i)\over L(1)}~|~n_i\in{\bf Z},~~
\alpha_i\in{\overline {\bf Q}}\cap {\bf R}~~{\rm for~all}~~
i\right \} \cap{\bf Q}.
\eqno (1.6)
$$

According to a conjecture of ${\cal W}$. Nahm [Na] the abelian group
${\cal W}$ ``coincides''
with the spectra in rational conformal field theory.
Thus it seems very interesting
task  to obtain more explicit description of the group ${\cal W}$
(e.g. to find a system of generators for ${\cal W}$) and also to connect
already known results about
the spectra in conformal field theory (see e.g. [BPZ], [FF], [GKO], [FQS],
[Bi], [Ka], [KP]) with suitable
elements in ${\cal W}$. One of our main results of the present paper allows to
describe some part of a system of generators for abelian group ${\cal W}$.
As a corollary, we will show that the spectra of unitary minimal models
[BPZ], [GKO] and some others are really contained in ${\cal W}$. But at
first we remind some already known
relations between the values of the Rogers dilogarithm function.

\vskip 1cm

{\bf \S 2.~~Some dilogarithm relations.}

\bigbreak

It is easy to see from (1.2) and (1.3) that
$$ L({1\over 2})={\pi^2\over 12},~~L({1\over 2}(\sqrt 5-1))={\pi^2\over 10},~~
L({1\over 2}(3-\sqrt 5))={\pi^2\over 15}.\eqno (2.1)
$$
Proof.~~Let us put $\alpha :={1\over 2}(\sqrt 5-1)$. It is clear that
$\alpha^2+\alpha =1$. So we have $L(\alpha^2)=L(1-\alpha)=L(1)-L(\alpha)$.
Now we use the Abel's formula
$$L(x^2)=2L(x)-2L\left({x\over 1+x}\right), \eqno (2.2)
$$
which may be obtained from (1.3) in the case $x=y$. From (2.2) we find that
$$L(\alpha^2)=2L(\alpha )-2L({\alpha\over 1+\alpha })=
2L(\alpha )-2L(\alpha ^2).$$
So, $3L(\alpha ^2)=2L(\alpha )$. But as we already saw,
$$L(1)=L(\alpha )+L(\alpha ^2)={5\over 3}L(\alpha ).$$
This proves (2.1).

\qed

Apparantly, there are no other algebraic point from the interval $(0,1)$ at
which there is such an elementary evaluation of Rogers dilogarithm function.
However, there are many identities relating the values of dilogarithm
function at various powers of algebraic numbers. It is interesting to write
some of this identities in order to compare the elements of the abelian group
$W$ obtained by such manner with the spectra of known conformal models. As
concerning identities between the values of dilogarithm function, we follow
[Le1], [Le2], [Lo] and [RS].
$$ \eqalignno{
&6L({1\over 3})-L({1\over 9})={\pi^2\over 3}, & (2.3)\cr
&\sum_{k=2}^nL\left({1\over {k^2}}\right)+2L\left({1\over n+1}\right) =
{\pi^2\over 6}, & (2.4)\cr }
$$
consequently,
$$\sum_{k=2}^\infty L\left({1\over k^2}\right) ={\pi^2\over 6}.$$

This identities may be easily deduced from (1.2) and (2.2). Again, if
$\alpha =\sqrt 2-1$, we have the relations
$$\eqalignno{
&4L(\alpha )+L(1-\alpha^2)={5\pi^2\over 12}, \cr
&4L(\alpha )+4L(\alpha^2)+L(1-\alpha^4)={7\pi^2\over 12} & (2.5)\cr }
$$
Proof.~ We use Abel's formula (2.2) and relation (1.3) with $y={1\over 2}$
$$ L({1\over 2})+L(x)=L\left({x\over 2}\right)+L\left({x\over 2-x}\right)+
L\left({1-x\over 2-x}\right).\eqno (2.6)$$
We have
$$\eqalignno{
&L(\alpha^2)=2L(\alpha )-2L\left( {\alpha\over 1+\alpha}\right)=2L(\alpha)-
2L\left( {2-\sqrt 2\over 2}\right), \cr
&L({1\over 2})=2L({\sqrt 2\over 2})-2L(\alpha )={\pi^2\over 3}-2L({2-\sqrt 2
\over 2})-2L(\alpha ), \cr
&L(\alpha^4)=2L(\alpha^2)-2L\left({\alpha^2\over 1+\alpha^2}\right)=
2L(\alpha^2)-2L\left({2-\sqrt 2\over 4}\right),\cr
&L\left({2-\sqrt 2\over 2}\right)+L({1\over 2})=
L\left({2-\sqrt 2\over 4}\right)+L(\alpha )+L(\alpha^2).\cr }
$$
Excluding successively $L({2-\sqrt 2\over 2})$ and $L({2-\sqrt 2\over 4})$
from these relations we obtain (2.5).

\qed

Watson [Wa] found three relations involving the roots of the cubic
$x^3+2x^2-x-1$. Namely, if we take $\alpha =
\displaystyle{{1\over 2}\sec{2\pi\over 7}},~~
\beta =\displaystyle{{1\over 2}\sec{\pi\over 7}},~~
\gamma =2\displaystyle{\cos{3\pi\over 7}}$,
then $\alpha$, $-\beta$ and $-\displaystyle{{1\over\gamma}}$
are the roots of this cubic and
$$\eqalignno{
&L(\alpha )+L(1-\alpha^2)={4\pi^2\over 21},\cr
&2L(\beta )+L(\beta^2)={5\pi^2\over 21}, & (2.7)\cr
&2L(\gamma )+L(\gamma^2)={4\pi^2\over 21}.}
$$
%Proof.~
%
%
%
%

%\qed

Lewin [Le1] and Loxton [Lo] found three relations involving the roots of the
cubic $x^3+3x^2-1$. Namely, if we take $\delta =
\displaystyle{{1\over 2}\sec{\pi\over 9}}$,
$\epsilon =\displaystyle{{1\over 2}\sec{2\pi\over 9}}$,
$\zeta =2\displaystyle{\cos{4\pi\over 9}}$, then
$\delta,~~-\epsilon$ and $-\displaystyle{{1\over\zeta}}$
are the roots of this cubic and
$$\eqalignno{
&3L(\delta )+3L(\delta^2)+L(1-\delta^3)={17\pi^2\over 18},\cr
&6L(\epsilon )+9L(1-\epsilon^2)+2L(1-\epsilon^3)+L(\epsilon^6)=
{31\pi^2\over 18},&(2.8)\cr
&6L(\zeta )+9L(1-\zeta^2)+2L(1-\zeta^3)+L(\zeta^6)={35\pi^2\over 18}.}
$$
\vskip 1cm

{\bf \S 3.~~Basic~identities~and~conformal~weights.}

\bigbreak

In this section we present our main results dealing with a computation of the
following dilogarithm sum
$$
\sum_{k=1}^{n-1}\sum_{m=1}^rL\left({\sin k\varphi\cdot
\sin (n-k)\varphi\over\sin
 (m+k)\varphi\cdot\sin (m+n-k)\varphi}\right)
%={(n^2-1)r\over r+n}\cdot {\pi^2\over 6}
:={\pi^2\over 6}s(j,n,r), \eqno (3.1)
$$
 where $ \varphi =\displaystyle{{(j+1)\pi\over n+r}},~~0\le j\le n+r-2$.

It is clear that $s(j,n,r)=s(n+r-2-j,n,r)$, so we will assume in sequel that
$0\le 2j\le n+r-2$.

The dilogarithm sum (3.1) corresponds to the Lie algebra of type $A_{n-1}$.
The case $j=0$ was considered in our previous paper [Ki1], where it was proved
that $s(0,n,r)=\displaystyle{{(n^2-1)r\over n+r}}$.
It was stated in [Ki1] that this number
coincides with the central charge of the $SU(n)$ level $r$ WZNW model.
Before formulating our result about computation of the sum $s(j,n,r)$ let us
remind the definition of Bernoulli polynomials. They are defined by the
generating function
$${te^{xt}\over e^t-1}=\sum_{n=0}^{\infty}B_n(x){t^n\over n!},~~|t|<2\pi.
$$
We use also modified Bernoulli polynomials
$$ {\overline B}_n(x)=B_n(\{ x\} ),~~{\rm where}~~\{ x\} =x-[x]
$$
be a fractional part of $x\in {\bf R}$. It is well-known that
$$\eqalignno{
&{\overline B}_{2n}(x)=(-1)^n{2(2n)!\over (2\pi )^{2n}}\sum_{k=1}^{\infty }
{\cos 2k\pi x\over k^{2n}},\cr
&    &(3.2)\cr
&{\overline B}_{2n+1}(x)=(-1)^{n-1}{2(2n+1)!\over (2\pi )^{2n+1}}
\sum_{k=1}^{\infty }{\sin 2k\pi x\over k^{2n+1}}.}
$$

\medbreak

{\bf Theorem~3.1.}~~We have
$$s(j,n,r)=6(r+n) \sum_{k=0}^{[{n-1\over 2}]}\left\{
{1\over 6}-{\overline B}_2\biggl( (n-1-2k)
\theta\biggr) \right\} -{1\over 4}\biggl\{ 2n^2+1+3(-1)^n\biggr\},\eqno (3.3)
$$
where $\theta =\displaystyle{{j+1\over r+n}}$ and ${\rm g.c.d.}(j+1,r+n)=1$.

\medbreak

{\bf Theorem~3.2.}~(level-rank duality [SA], [KN1])
$$s(j,n,r)+s(j,r,n)=nr-1. \eqno (3.4)
$$

\medbreak

{\bf Corollary~3.3.}~~We have
$$s(j,n,r)=c_r^{(n)}-24h_j^{(r,n)}+6\cdot{\bf Z}_+, \eqno (3.5)
$$
where
$$c_r^{(n)}={(n^2-1)r\over n+r},~~h_j^{(r,n)}=
{n(n^2-1)\over 24}\cdot{j(j+2)\over r+n},~~0\le j\le r+n-2,\eqno (3.6)
$$
are the central charge and conformal dimensions of the $SU(n)$ level $r$ WZNW
primary fields, respectively.

Proof.~~Let us remind that
$$B_2(x)=x^2-x+{1\over 6}~~{\rm and}~~{\overline B}_2(x)=B_2(\{ x\} ). $$
Thus,
$$\eqalignno{
&6(r+n)\sum_{k=0}^{[{n-1\over 2}]}\left({1\over 6}-{\overline B}_2\biggl(
(n-1-2k)\theta\biggr) \right) =\cr
&=6(r+n)\sum_{k=0}^{[{n-1\over 2}]}(n-1-2k)\theta -6(r+n)\sum_{k=0}^{[{n-1
\over 2}]}(n-1-2k)^2\theta ^2+\cr
&+6(r+n)\sum_{k=0}^{[{n-1\over 2}]}
\biggr[(n-1-2k)\theta\biggl]\biggl((n-1-2k)\theta-1+\biggl\{ (n-1-2k)\theta
\biggr\}\biggr):=\cr
&~~\cr
& =6\Sigma_1-6\Sigma_2+6\Sigma_3. }
$$
Now if we take $\theta =\displaystyle{{j+1\over r+n}}$, then  it is clear that
$\Sigma_3\in{\bf Z}_+$. In order to compute $\Sigma_1$ and $\Sigma_2$ we use
the following summation formulae
$$\eqalignno{
&\sum_{k=0}^{[{n-1\over 2}]}(n-1-2k)={2n^2-1+(-1)^n\over 8}=
\left[{n\over 2}\right]\cdot\left[ {n+1\over 2}\right],\cr
&\sum_{k=0}^{[{n-1\over 2}]}(n-1-2k)^2={n(n^2-1)\over 6}.}
$$
Consequently ~~ $s(j,n,r)=$
$$\eqalignno{
&={3(2n^2-1+(-1)^n)(j+1)\over 4}-{n(n^2-1)(j+1)^2\over r+n}-{2n^2+1+
3(-1)^n\over 4}+6\Sigma_3=\cr
&={(n^2-1)r\over r+n}-{n(n^2-1)j(j+2)\over r+n}+6j
\left[{n\over 2}\right]\cdot\left[ {n+1\over 2}\right] +6\Sigma_3.}
$$
\qed

For small values of $j$ we may compute the sum in (3.3) and thus to find
corresponding positive integer in (3.5).

\medbreak

{\bf Corollary~3.4.}

$i)$ if ~$j\le r$,~ then
$$s(j,2,r)=c_r^{(2)}-24h_j^{(r,2)}+6j={3r\over r+2}+6~{j(r-j)\over r+2}.
\eqno (3.7)
$$

$ii)$ if ~$2j\le r+1$, ~ then
$$s(j,3,r)={8r\over r+3}-24~{j(j+2)\over r+3}+12j. \eqno (3.8)
$$

$iii)$ if ~$(n-1)j<r+1$,~ then
$$s(j,n,r)=c_r^{(n)}-24h_j^{(r,n)}+6j\cdot \left[{n\over 2}\right]
\cdot\left[{n+1\over 2}\right]. \eqno (3.9)
$$

Proof.~~An assumption ~$(n-1)j<r+1$~ is equivalent to a condition\break
$\displaystyle{{(n-1)(j+1)\over n+r}}<1$. So the term $\Sigma_3$
(see a proof of Corollary 3.3) is equal to zero.

\qed

It seems interesting to find a meaning of the positive integer in (3.5).
Now we want to find a ``dilogarithm interpretation'' of the central charges
and conformal dimensions for some well-known conformal models.

\medbreak

{\bf Corollary~3.5.}~~We have
$$s(j_1,2,k)+s(0,2,1)-s(j_2,2,k+1)=c_k-24h_{j_1+1,j_2+1}+6(j_1-j_2)(j_1-j_2+1),
\eqno (3.10)
$$
where
$$c_k=1-{6\over (k+2)(k+3)},~~h_{r,s}^{(k)}={[(k+3)r-(k+2)s]^2-1
\over 4(k+2)(k+3)} \eqno (3.11)
$$
are the central charge and conformal dimensions of the primary fields for
unitary minimal conformal models [BPZ], [Ka], [GKO].

\medbreak

{\bf Corollary~3.6.}
$$s(j_1,n,k)+s(0,n,1)-s(j_2,n,k+1)=c_{k,n}-24h_{j_1+1,j_2+1}^{(n)}+12{\bf Z}_+,
\eqno (3.12)
$$
where
$$\eqalignno{
&c_{k,n}=(n-1)\left\{ 1-{n(n+1)\over (k+n)(k+n+1)}\right\} ,\cr
&~~ &(3.13)\cr
&h_{r,s}^{(n)}(k)={n(n^2-1)\over 24}\cdot{[(k+n+1)r-(k+n)s]^2-1\over
(k+n)(k+n+1)}}
$$
are the central charge and conformal dimensions of the primary fields for $W_n$
models [Bi], [CR].

\medbreak

{\bf Corollary~3.7.}
$$\eqalignno{
&s(j_1,2,k)+s\biggl({1\over 2}(1-(-1)^{j_1-j_2}),2,2\biggr)-
s(j_2,2,k+2)=\cr
& ~~&(3.14)\cr
&=c(k)-24{\widetilde h}_{j_1+1,j_2+1}+12\left[{j_1-j_2+1\over 2}\right]
\cdot\left[ {j_1-j_2+2\over 2}\right],}
$$
where
$$\eqalignno{
&c(k)={3\over 2}\left( 1-{8\over (k+2)(k+4)}\right),\cr
&  &(3.15)\cr
&{\widetilde h}_{r,s}:={\widetilde h}_{r,s}(k)={[(k+4)r-(k+2)s]^2-4\over
8(k+2)(k+4)}+{1-(-1)^{r-s}\over 32},}
$$
are the central charge and conformal dimensions of the primary fields for
unitary minimal $N=1$ superconformal models [GKO], [MSW].

We give a generalization of Corollaries (3.5)-(3.7) to the case of non-unitary
minimal models.

\medbreak

{\bf Corollary~3.8.}~~If $p\ge q\ge 2$, then
$$\eqalignno{
&(p-q)s(j_1,2,q-2)+s(0,2,1)-(p-q)s(j_2,2,p-2)=\cr
&  &(3.16)\cr
&=c-24h_{j_1+1,j_2+1}+
6(j_1-j_2)(p-q+j_1-j_2), }
$$
where
$$\eqalignno{
&c=1-{6(p-q)^2\over pq},& (3.17) \cr
&h_{r,s}:=h_{r,s}(c)={(pr-qs)^2-(p-q)^2\over 4pq},~~r< q,~s< p,}
$$
are the central charge and conformal dimensions of the primary fields for
non-unitary (if $p-q\ge 2$) Virasoro minimal models [FQS]. Note that
``remainder
term'' in (3.16)
$$6(j_1-j_2)(p-q+j_1-j_2)$$
appears to be positive for all $0\le j_1<q,~~0\le j_2<p$ iff $p-q=0$ (trivial
case) or $p-q=1$ (unitary case).

\medbreak

{\bf Corollary~3.9.}~~Let $p\ge q\ge 2$ and $p-q\equiv 0~~({\rm mod}~ 2)$. Then
$$ \eqalignno{
&{p-q\over 2}s(j_1,2,q-2)+s\biggl({1\over 2}(1-(-1)^{j_1-j_2}),2,2\biggr)
-{p-q\over 2}s(j_2,2,p-2)=\cr
&  &(3.18)\cr
&={\widetilde c}-24{\widetilde h}_{j_1+1,j_2+1}+6\left\{
{(j_1-j_2)(j_1-j_2+p-q)\over 2}+{1-(-1)^{j_1-j_2}\over 4}\right\},}
$$
where
$$\eqalignno{
&{\widetilde c}={3\over 2}\left( 1-{2(p-q)^2\over pq}\right),\cr
&~~ & (3.19)\cr
&{\widetilde h}_{r,s}={(pr-qs)^2-(p-q)^2\over 8pq}+{1-(-1)^{r-s}\over 32},
{}~~r<q,~s<p,}
$$
are the central charge and conformal dimensions of primary fields for
non-unitary  (if $p-q>2$) Neveu-Schwarz (if $r-s$ even) or Ramond
(if $r-s$ odd) minimal models [FQS].

\medbreak

{\bf Corollary~3.10.}~~ Let $p\ge q\ge n$, then
$$\eqalignno{
&(p-q)s(j_1,n,q-n)+s(0,n,1)-(p-q)s(j_2,n,p-n)=\cr
&~~&(3.20)\cr
&=c-24h_{j_1+1,j_2+1}(c)+6{\bf Z},}
$$
where
$$\eqalignno{
&c=(n-1)\left\{ 1-{n(n+1)(p-q)^2\over pq}\right\} ,\cr
&  &(3.21)\cr
&h_{rs}(c)={n(n^2-1)\over 24}{(pr-qs)^2-(p-q)^2\over pq},~~r<q,~s<p,}
$$
are the central charge and conformal dimensions of primary fields for
non-unitary (if $p-q\ge 2$) $W_n$ minimal models [Bi].

Finally we give a ``dilogarithm interpretation'' for the central charges and
conformal weights of restricted solid-on-solid (RSOS) lattice models and their
fusion hierarchies [KP].

\medbreak

{\bf Corollary~3.11.}~~We have
$$\eqalignno{
&s(l,2,N)+s(N-1,2,N-2)-s(m-1,2,N-2)= &(3.22)\cr
&=c+1-24\Delta +6(l-|m|),~~m\in{\bf Z},~~
0\le l\le N, }
$$
where
$$c={2(N-1)\over N+2}~~{\rm and}~~\Delta ={l(l+2)\over 4(N+2)}-{m^2\over 4N}
\eqno (3.23)
$$
are the central charge and conformal weights of ${\bf Z}_N$ parafermion
theories  [FZ]. The members of (3.23) may be also realized as the central
charge and conformal weights of fusion $N+1$-state RSOS($p,p$) lattice models
[DJKMO], [BR] on the regime I/II critical line. Note that physical constraints
$$|m|\le l,~~m\equiv l~({\rm mod}~2)$$
for value of $m$ in (3.23) are equivalent to a condition that ``remainder
term'' in (3.22), namely, $6(l-|m|)$, must be belongs to $12{\bf Z}_+$.

\medbreak

{\bf Corollary~3.12.}~~Let us fix the positive integers $k,~~p=1,2,\ldots$
(the fusion level), $j_1$ and $j_2$ such that
$0\le j_1\le k,~~0\le j_2\le k+l$.

Let $r_0=p\left\{ \displaystyle{{j_1-j_2\over p}}\right\}$
be the unique interger determined by
$$0\le r_0\le p,~~r_0\equiv\pm (j_1-j_2){\rm mod} ~2p. \eqno (3.24)
$$
Then we have
$$\eqalignno{
&s(j_1,2,k)+s(r_0,2,p)-s(j_2,2,k+p)=& (3.25)\cr
&~~~ \cr
&=c-24\Delta +12~{(j_1-j_2)(p+j_1-j_2)
+r_0(p-r_0)\over 2p},}
$$
where
$$\eqalignno{
&c={3p\over p+2}\left( 1-{2(p+2)\over (k+2)(k+p+2)}\right), \cr
&  &(3.26)\cr
&\Delta ={[(k+p+2)(j_1+1)-(k+2)(j_2+1)]^2-p^2\over 4p(k+2)(k+p+2)}+
{r_0(p-r_0)\over 2p(p+2)}}
$$
are the central charge and conformal weights of the fusion $(k+p+1)$-state
RSOS($p,p$) latice models [KP] on the regime III/IV critical line. It is easy
to see that ``remainder term'' in (3.25) belongs to $12{\bf Z}_+$. Note also
that the fusion RSOS($p,q$) lattice models, obtained by fusing~ $p\times q$~
blocks of face weights together, are related to coset conformal fields
theories obtained by the Goddard-Kent-Olive (GKO) construction [GKO]. Namely,
$c$ and $\Delta$ in (3.26) are the central charge and conformal dimensions
of conformal field theory, which corresponds to the coset pair [GKO]
$$\matrix{
&&A_1&\oplus& A_1&\supset& A_1\cr
{\rm levels}&&k&&p&&k+p\cr}
$$
Thus the members of (3.26) are
reduced to those of (3.11) if $p=1$ and of (3.15) if $p=2$.

\vskip 1cm

{\bf \S 4.~~$A_1$-type~ dilogarithm~ identities.}

\bigbreak

As is well-known [Le2], the Rogers dilogarithm function $L(x)$ admits a
continuation on all complex plane ${\bf C}$. Follow [Le], [KR] we define
a function
$$\eqalignno{
L(x,\theta ):&=
-{1\over 2}\int_0^x{\log (1-2x\cos\theta+
x^2)\over x}dx+{1\over 4}\log |x|\cdot\log (1-2x\cos\theta +x^2)=\cr
&=ReL(xe{^{i\theta}}),~~x,\theta\in\bf R & (4.1) }
$$
Our proof of Theorem 3.1 is based on a study of properties of the function
$L(x,\theta )$.

\medbreak

{\bf Proposition~4.1.}~~ For all real $\varphi,~\theta$ we have
$$\eqalignno{
L\Big (\Big ({\sin\theta\over\sin\varphi}\Big )^2\Big )&=
%2\varphi\cdot\theta
\pi^2\biggl\{{\overline B}_2\left({\theta+\varphi\over\pi}\right) -
{\overline B}_2\left({\varphi\over\pi}
\right) -{\overline B}_2\left({\theta\over\pi}\right) +{1\over 6}\biggr\}+\cr
&~~&(4.2)\cr
&+2L\Big (-{\sin (\varphi -\theta )\over\sin\theta},\varphi\Big )-
2L\Big (-{\sin\varphi\over\sin\theta},\varphi+\theta\Big ). \cr}
$$
Before proving a Proposition 4.1 let us give the others useful properties
of function (4.1) (compare with [Le2]).

\medbreak

{\bf Lemma~4.2.}
$$\eqalignno{
(i)~~~~&L(x,0)=L(x),~~L(-x,\varphi )=L(-x,\pi -\varphi ) & (4.3) \cr
(ii)~~~&L(x,\varphi )=L(x,2\pi k\pm\varphi ),~~k\in {\bf Z } & (4.4) \cr
(iii)~~&L(-1,\varphi)=\pi^2{\overline B}_2\left({\varphi\over 2\pi}+
{1\over 2}\right), \cr
&L(1,\varphi)=\pi^2{\overline B}_2\left({\varphi\over 2\pi}\right) & (4.5) \cr
(iv)~~~&L(x,\varphi )+L(x^{-1},\varphi )=2\pi^2{\overline B}_2\left({\varphi
\over
2\pi}\right),
{}~~x>0 \cr
&L(-x,\varphi )+L(-x^{-1},\varphi )=
2\pi^2{\overline B}_2\left({\varphi\over 2\pi}+{1\over 2}\right),
{}~~x<0~~~~~~~~~~~~~~~~~~~~~~~~& (4.6) \cr
(v)~~~~&L(0,\varphi)=0,~~L(+\infty ,\varphi )=
2\pi^2{\overline B}_2\left({\varphi\over 2\pi}
\right), \cr
&L(-\infty ,\varphi )=2\pi^2{\overline B}_2\left({\varphi +\pi\over
2\pi}\right)
& (4.7) \cr
(vi)~~~&L(2\cos\varphi ,\varphi )=\pi^2\biggl\{ {\overline B}_2\left({\varphi
\over \pi}\right) +{1\over 12}\biggr\} & (4.8) \cr
(vii)~~&L(x^n,n\varphi)=n\sum_{k=0}^{n-1}L\biggl(x,~\varphi +
{2k\pi\over n}\biggr),~~
x\in{\bf R}_+ ,\cr
&L(x^n)=n\sum_{k=0}^{n-1}L\left( x\cdot \exp{2k\pi i\over n}\right)
,~~x\in (0,1).
& (4.9) \cr}
$$
More generally (Rogers' identity [Ro])
$$L(1-y^n)=\sum_{k=1}^n\sum_{l=1}^n\biggl[ L(\lambda_k/\lambda_l)-
L(x_k\lambda_l)\biggr] ,$$
where $\{ x_k\}^n_{k=1}$ are the roots of the equation
$$ 1-y^n=\prod^n_{k=1}(1-\lambda_kx).$$

Proof.~~At first let us remind some properties of modified Bernoulli
polynomials.
$$\eqalignno{
&{d{\overline B}_n(x)\over dx}=n{\overline B}_{n-1}(x), \cr
&{\overline B}_n(x)={\overline B}_n(x+1),~~{\overline B}_n(-x)=
(-1)^n{\overline B}_n(x),
&(4.10) \cr
&{\overline B}_p(nx)=n\sum^n_{k=1}{\overline B}_p\biggl(x+{k\over n}\biggr).}
$$

Note that identities (4.5) follows from the Fourier expansion for
${\overline B}_2(x)$ (see (3.2)).

In order to prove the identity (4.2) let us differentiate LHS and RHS of the
last one with respect to $\varphi$ using the following formula
$$\eqalignno{
&dL(x,\varphi )=\left\{ -{1\over 4}~{\log (1-2x\cos\varphi +x^2)\over x}+
{1\over 2}\log |x|{x-\cos \varphi\over 1-2x\cos\varphi+x^2}\right\}dx \cr
&~~&(4.11)\cr
&+\left\{ -\tan^{-1}\left({x\sin\varphi\over 1-x\cos\varphi}\right) +
{1\over 2}\log |x|{x\sin\varphi\over 1-2x\cos\varphi +x^2}\right\}d\varphi .}
$$
Acting in such a manner we find
$$\eqalignno{
&{d\over d\varphi }L\left(-{\sin\varphi\over\sin\theta},\varphi +\theta\right)
=\cr &~~~~~~~~~~~~=\varphi +{1\over 2}{\rm cot}(\varphi +\theta )\log \left(
{\sin\varphi\over\sin\theta
}\right) -{1\over 2}{\rm cot}\varphi\log\left({\sin (\varphi +\theta )
\over\sin\theta}
\right),~~~~&(4.12)\cr
&{d\over d\varphi }L\left(-{\sin\theta\over\sin\varphi},\varphi +\theta\right)=
\cr &~~~~~~~~~~~~=
\theta-{1\over 2}{\rm cot}(\varphi +\theta )\log \left({\sin\varphi
\over\sin\theta}\right) -{1\over 2}{\rm cot}\varphi\log\left({\sin (\varphi
+\theta )\over\sin\theta}\right),~~~~&(4.13)\cr
&{\rm if}~~0<\varphi ,~~\theta<\pi,~~\varphi +\theta <\pi ;&(4.13a)}
$$
$$\eqalignno{
&{d\over d\varphi }L\Bigg ({\sin\theta\over\sin\varphi}\cdot{\sin (\theta
+\psi )\over\sin(\varphi+\psi )}\Bigg )=\cr
&~~~~~~~~~~~~~={1\over 2}\biggl[{\rm cot}\varphi +
{\rm cot}(\varphi +\psi )\biggr]
\log \left({\sin (\varphi -\theta )\sin (\varphi +\theta +\psi )
\over\sin\theta\sin (\theta +\psi)}\right) -~~~~\cr
&~~~~~~~~~~~~-{1\over 2}\biggl[{\rm cot}(\varphi -\theta )+{\rm cot}(\varphi +
\theta +\psi )\biggr]\log\left({\sin\varphi\sin (\varphi +\psi )
\over\sin\theta\sin (\theta +\psi )}
\right),~~~~&(4.14)\cr
&      \cr
&{\rm if}~~0<\theta <\varphi <\pi ,~~0<\psi <\pi,~~\varphi +\theta +\psi <\pi .
&(4.14a)}
$$

Further, let us use the reduction rules (4.3), (4.4) and (4.6)
(compare with (1.4)
and (1.5)) if the angles $\varphi,~\theta$~ (or $\varphi ,~\theta ,~\psi $)
do not satisfy the condition (4.13a) (or (4.14a)). As a result one can obtain
that a derivative of difference between LHS and RHS of (4.2) with respect
to $\varphi$ is equal to
$$ 2\pi\left\{ {\overline B}_1\left({\varphi +\theta\over \pi}\right) -
{\overline B}_1\left({\varphi\over\pi}\right)\right\}. $$
Integrating (see (4.10)), then taking $\varphi =0$ and using (4.5) to
determine the integration constant, we obtain the equality (4.2).

It is easy to see that (4.8) follows from (4.2) when $\varphi +\theta =\pi$. In
order to prove (4.9) let us differentiate LHS and RHS of (4.9) with respect
to $x$ and use a summation formula
$$\sum_{k=0}^{n-1}{\exp i(\varphi +{2\pi k\over n})\over 1-x\exp i(\varphi +
{2\pi k\over n})}={nx^{n-1}\exp (in\varphi )\over 1-x^n\exp (in\varphi )}.
$$
Thus, the proofs of Proposition 4.1 and Lemma 4.2 are finished.

\qed

Proof of the Theorem 3.1 for the case $n=2$.~~
If we substitute $\theta=m\varphi $ in (4.2) then obtain
$$\eqalignno{
&L\Bigg (\left({\sin m\varphi\over\sin\varphi}\right)^2\Bigg )=
\pi^2\biggl\{ {\overline B}_2\biggl({(m+1)\varphi\over\pi}\biggr)-
{\overline B}_2\biggl({m \varphi\over \pi}\biggr)-
{\overline B}_2\biggl({\varphi\over \pi}\biggr)+{1\over 6}\biggr\} +\cr
&+2L\biggl(-{\sin ((m-1)\varphi )\over\sin\varphi },m\varphi\biggr)-
2L\biggl(-{\sin m\varphi \over \sin\varphi},(m+1)\varphi \biggr) .& (4.15)}
$$
Futher let us introduce notation
$$ f_m(\varphi ):=1-{Q_{m-1}(\varphi )Q_{m+1}(\varphi )\over Q^2_m(\varphi )}
={1\over Q_m^2(\varphi )}.$$
Then using (4.15) we find
$$\eqalignno{
&\sum^r_{m=1}L(f_m(\varphi ))=
-2\biggl\{ L\biggl( -Q_r(\varphi ),(r+2)\varphi \biggr)
-{\pi^2\over 6}\biggr\}+ &(4.16)\cr
&+\pi^2\left\{ {\overline B}_2\left({(r+2)\varphi\over\pi}\right)
-{1\over 6}\right\} +(r+2)\pi^2\left\{ {1\over 6} -
{\overline B}_2\left({\varphi\over\pi}\right)\right\} -
{\pi^2\over 2}.}
$$
Now let us put $\varphi =\displaystyle{{(j+1)\pi\over r+2}},
{}~~0\le j\le r+1$. Then
$Q_r(\varphi )=(-1)^j$ and it is clear from (4.3) and (4.4) that
$$ L\biggl( (-1)^{j+1},~(j+1)\pi \biggr) =L(1)={\pi^2\over 6}.$$
\qed

Note that polynomials $Q_m:=Q_m(\varphi )$ satisfy the following recurrence
relation
$$Q_m^2=Q_{m-1}Q_{m+1}+1,~~Q_0\equiv 1,~~m\ge 1,$$
where as the polynomials $y_m:=y_m(\varphi )=
Q_{m-1}(\varphi )\cdot Q_{m+1}(\varphi )$ satisfy the following one
$$ y_m^2=(1+y_{m-1})(1+y_{m+1}),~~y_0\equiv 0,~~m\ge 1.$$

Follow [Le2] we define a function $W(x,\varphi )$ by
$$\eqalignno{
&W(x,\varphi ):=W(x,\varphi ,\theta )=L\left({\sin ^2\theta\over\sin\varphi
(\sin\varphi +x\sin (\varphi +\theta ))}\right)+\cr
&~~~&(4.17)\cr
&+L\left( -{x^2\sin\varphi +x\sin (\varphi -\theta )\over x\sin (\varphi +
\theta )+\sin\varphi}\right) -L\left({x\sin (\varphi +\theta )\over x\sin
(\varphi +\theta )+\sin\varphi}\right) .}
$$
Note the following particular cases
$$\eqalignno{
&W(0,\varphi )=L\left(\left({\sin\theta\over\sin\varphi }\right)^2\right),~~
W(-2\cos\theta ,\varphi )=L\left(-{\sin^2\theta\over\sin\varphi\sin (\varphi +
2\theta )}\right),& (4.18)\cr
&W(1,\varphi )=L\left({\sin\theta\sin{1\over 2}\theta\over\sin\varphi
\sin (\varphi +{1\over 2}\theta )}\right)
+L\left({\sin ({1\over 2}\theta -\varphi )\over\sin ({1\over 2}\theta
+\varphi )}\right) -L\left({\sin (\varphi +\theta )
\over 2\sin (\varphi +{1\over 2}\theta )\cos{1\over 2}\theta }\right),\cr
&~~\cr
&W(-1,\varphi ,\theta )=W(1,\varphi ,\pi +\theta ).}
$$
\medbreak

{\bf Proposition 4.3.}~~We have
$$\eqalignno{
W(x,\varphi )&=2L(-x,\theta )+2L(-x_1,\varphi )-2L(-x_2,\varphi +\theta )+
{}~~~~~~~~~~~~~~~~\cr
&~~\cr
&+\pi^2\left\{ {\overline B}_2\left({\varphi +\theta\over\pi}\right) -
{\overline B}_2\left({\varphi\over\pi}\right) -
{\overline B}_2\left({\theta\over\pi}\right) +
{1\over 6}\right\} ,&(4.19)\cr
&~~\cr
&~~~\cr
{\rm where}~~~~~~~~x_1=&{x\sin\varphi +
\sin (\varphi -\theta )\over\sin\theta}~~
{\rm and}~~x_2={x\sin (\varphi +\theta )+\sin\varphi\over\sin\theta }.}
$$
A proof of Proposition 4.3 may be obtained by the same manner as that of
Proposition 4.1.

\qed

Note that $x_2$ is obtained from $x_1$ by replacing $\varphi$ by
$\varphi +\theta$. The angular parameter in (4.19) also increases in this way
and the terms $L(-x,\varphi )$ and $L(-x_2,\varphi +\theta )$ have opposite
signs.  So if we substitute successively the angles $\varphi ,\varphi +\theta,
\ldots ,\varphi +r\theta$ instead of $\varphi$ in (4.19) and after this will
add all results together, we will obtain the following generalisation of
(4.16):
$$\eqalignno{
&\sum^r_{k=0}W(x,\varphi +k\theta )=2(r+1)L(-x,\theta )+2L(-x_1,\varphi )-
2L(-x_{r+2},\varphi +(r+1)\theta )+\cr
& +\pi^2\left\{ {\overline B}_2\left( {\varphi +(r+1)
\theta\over\pi}\right) -{\overline B}_2\left({\varphi\over\pi}\right)\right\} +
(r+1)\pi^2\left\{ {1\over 6}-{\overline B}_2\left(
{\theta\over\pi}\right)\right\} ,
& (4.20)\cr
&~~~\cr
&{\rm where}~~~~x_{n+1}:={x\sin (\varphi + n\theta )+
\sin (\varphi +(n-1)\theta )
\over\sin\theta },~~0\le n\le r+2.}
$$
Now let us take $\varphi =\theta$ in (4.20).
Then we find $x_1=x$, so that (4.20)
becomes
$$\eqalignno{
\sum_{k=1}^{r+1}&W(x,k\theta )=2(r+2)L(-x,\theta )+(r+2)\pi^2\left\{{1\over 6}
-{\overline B}_2\left({\theta\over\pi }\right)\right\} -
{\pi^2\over 3}+~~~~~~~~~~~ &(4.21)\cr
&+\pi^2\left\{ {\overline B}_2\left({(r+2)\theta\over\pi}\right) -
{1\over 6}\right\}-2\left( L(-x_{r+2},(r+2)\theta )-{\pi^2\over 6}\right),\cr
& \cr
{\rm where}~~~~&x_{n+1}:=x_{n+1}(x,\theta )={x\sin(n+1)\theta +
\sin n\theta\over\sin\theta}.}
$$
Note the following particular cases ($0\le n\le r+2$)
$$\eqalignno{
&x_{n+1}(0,\theta )={\sin n\theta\over\sin\theta },~~x_{n+1}(-2\cos\theta ,
\theta )=-{\sin (n+2)\theta\over\sin\theta},\cr
&~~~\cr
&x_{n+1}(1,\theta )={\sin\left({1\over 2}(2n+1)\theta\right)
\over\sin{1\over 2}\theta },~~x_{n+1}(-1,\theta )=x_{n+1}(1,\theta +\pi ).}
$$
One can show that an identity (4.21) is reduced to (4.2) if $x=-2\cos\theta $
(or $x=0$). Now assume~ $x\ne -2\cos\theta$~ and take ~$\theta =
\displaystyle{{(j+1)\pi\over r+2}}$~ in (4.21).
Then we find ~$x_{r+2}=(-1)^j$, so that (4.21) becomes
$$\eqalignno{
&\sum_{k=1}^{r+1}W(x,k\theta )=2(r+2)L(-x,\theta )+{\pi^2\over 6}(1+s(j,2,r)),
{}~~~~~~~~~~~~~~&(4.22)\cr
&~~\cr
{\rm where}~~~~&s(j,2,r)={3r\over r+2}+
6~{j(r-j)\over r+2}~~({\rm see}~~(3.7)).}
$$

Finally let us take $x=\pm 1$ in (4.22). After some manipulations we obtain the
following result.

\medbreak

{\bf Proposition 4.4.}~~Let functions~ $W(\pm 1,\theta )$~
are defined by (4.18)
and\break
$\theta =\displaystyle{{(j+1)\pi\over r+2}}$. Then we have
$$\eqalignno{
&\sum_{k=1}^{r+1}W(-1,k\theta )={\pi^2\over 6}\left\{ 2r+2-{3(j+1)^2\over r+2}
\right\} ,\cr
&\sum_{k=1}^{r+1}W(+1,k\theta )={\pi^2\over 6}\left\{ 2-r-{3(j+1)^2\over r+2}
+6j\right\} .&(4.23)}
$$
\qed

Now we propose a generalisation of (4.16). Given a rational number
$p$ and decomposition of $p$ into the continued fraction
$$p=[b_r,b_{r-1},\ldots ,b_1,b_0]=b_r+{1\over\displaystyle b_{r-1}+
{\strut 1\over\displaystyle\cdots +{\strut 1\over\displaystyle b_1+
{\strut 1\over
\displaystyle b_0}}}}.\eqno (4.24)
$$
We will assume that $b_i>0$ if $0\le i<r$ and $b_r\in{\bf Z}$.
Using the decomposition (4.24) we define the set of integers $y_i$ and $m_i$:
$$\eqalignno{
&y_{-1}=0,~~y_0=1,~~y_1=b_0,\ldots ,y_{i+1}=y_{i-1}+b_iy_i,~~0\le i\le r,
&(4.25)\cr
&m_0=0,~~m_1=b_0,~~m_{i+1}=|b_i|+m_i,~~0\le i\le r.}
$$
It is clear that $p=\displaystyle{{y_{r+1}\over y_r}}$ and
$${y_{i+1}\over y_i}=p_i:=b_i+{1\over\displaystyle b_{i-1}+
{\strut 1\over\displaystyle\cdots +{\strut 1\over\displaystyle b_1+
{\strut 1\over
\displaystyle b_0}}}},~~0\le i\le r.
$$

The following sequences of integers were first introduced by Takahashi and
Suzuki [TS]
$$\eqalignno{
&r(j)=i,~~~{\rm if}~~m_i\le j<m_{i+1},~~0\le i\le r,\cr
&n_j=y_{i-1}+(j-m_i)y_i,~~{\rm if}~~m_i\le j<m_{i+1}+\delta_{i,r},~~0\le
i\le r.}
$$
Finally we define a dilogarithm sum of ``fractional level $p$'':
$$
\sum_{j=1}^{m_{r+1}}(-1)^{r(j)}L\Bigg (\left({\sin y_{r{(j)}}\theta\over
\sin (n_j+y_{r{(j)}})\theta}\right)^2\Bigg ):=
(-1)^r{\pi^2\over 6}s(k,2,p),\eqno (4.26)
$$
where $\theta =\displaystyle{{(k+1)\pi\over y_{r+1}+2y_r}}.$

The dilogarithm sum (4.26) (in the case $k=0$) was considered at first in
[KR], where its interpretation as a low-temperature asymptotic of the entropy
for the $XXZ$ Heisenberg model was given.
\medbreak

{\bf Proposition~4.5.}~~We have
$$\eqalignno{
& (i)~~~s(0,2,p)={3p\over p+2},\cr
&~~~&(4.27)\cr
& (ii)~~~s(k,2,p)={3p\over p+2}-6{k(k+2)\over p+2}+6{\bf Z}. }
$$
Proof. We start with
\medbreak

{\bf Lemma~4.6.}~~Given an integer $\sigma$ such that $m_i< \sigma\le m_{i+1}$.
Then for any $\theta\in{\bf R}$ we have
$$\eqalignno{
&\sum_{j=m_i}^{\sigma -1}L\Bigg (\left({\sin y_{r{(j)}}\theta\over
\sin (n_j+y_{r{(j)}})\theta}\right)^2\Bigg ):=
(\sigma-m_i)\pi^2\biggl\{ {1\over 6}-{\overline B}_2\biggl({y_i\theta
\over\pi}\biggr)\biggl\}+ \cr
&+\pi^2\biggl\{ {\overline B}_2\biggl({(n_{\sigma -1}
+2y_i)\theta\over \pi }\biggr)-{\overline B}_2\biggl({(y_{i-1}+y_i)\theta
\over \pi }\biggr)\biggr\} +&(4.28)\cr
&+2L\biggl(-{\sin y_{i-1}\theta\over\sin y_i\theta },(y_i+y_{i-1})\theta
\biggr)-2L\biggl(-{\sin (n_{\sigma -1}+y_i)\theta\over \sin y_i\theta},
(n_{\sigma -1}+2y_i)\theta\biggr) .}
$$
A proof of Lemma 4.6 follows from Proposition 4.1.

\qed

{}From (4.28) one can easily deduce the following generalisation of (4.16)
\medbreak

{\bf Corollary 4.7.}~~Given an integer $\sigma ,~~\sigma < m_i\le m_{i+1}$.
Then
$$\eqalignno{
&\sum_{j=0}^{\sigma -1}L\Bigg (\left({\sin y_{r{(j)}}\theta\over
\sin (n_j+y_{r{(j)}})\theta}\right)^2\Bigg ):=
\pi^2\sum^{i-1}_{j=0}(-1)^jb_j\left\{{1\over 6}-
{\overline B}_2\left({y_j\theta\over\pi}\right)\right\}+\cr
&+(-1)^i(\sigma -m_i)\pi^2\biggl\{ {1\over 6}-
{\overline B}_2\biggl({y_i\theta
\over\pi}\biggr)\biggl\}+(-1)^i\pi^2 {\overline B}_2\biggl(
{(n_{\sigma -1}+2y_i)\theta\over \pi }\biggr)-\cr
&-2\pi^2\sum^{i-1}_{j=0}(-1)^j{\overline B}_2\biggl({(y_{j-1}+y_j)\theta
\over \pi }\biggr)-
\pi^2{\overline B}_2\left({\theta\over\pi}\right)+&(4.29)\cr
&+2\sum_{j=0}^{i-1}(-1)^{j+1}\biggl\{ L\biggl(-{\sin y_{j-1}\theta
\over\sin y_j\theta },(y_j+y_{j-1})\theta
\biggr)+L\biggl(-{\sin y_j\theta\over\sin y_{j-1}\theta },(y_j+y_{j-1})\theta
\biggr)\biggr\}+\cr
&+(-1)^{i+1}2L\biggl(-{\sin (n_{\sigma -1}+y_i)\theta\over \sin y_i\theta},
(n_{\sigma -1}+2y_i)\theta\biggr) .}
$$

In order to go further we must compute the last sum in (4.29). Such
computation is based on the following result.
\medbreak

{\bf Lemma~4.8.}~~Given the real numbers $\varphi$ and $\theta$, let us define
$$\epsilon (\varphi ,\theta )=\{ {\varphi\over 2\pi}+{1\over 2}\} +
\{ {\theta\over 2\pi} +{1\over 2}\} -\{ {\varphi\over 2\pi }\} -
\{ {\theta\over 2\pi}\}~~~({\rm mod}~2), \eqno (4.30)
$$
where $\{ x\} =x-[x]$ be a fractional part of $x\in {\bf R}$.

Then we have
$$L\left(-{\sin\varphi\over\sin\theta},\varphi +\theta\right)+L\left(-{\sin
\theta\over\sin\varphi},\varphi+\theta\right)=
2\pi^2{\overline B}_2\left({\varphi+\theta +\epsilon (\varphi ,\theta )\pi
\over2\pi}\right) .\eqno (4.31)$$
A proof of Lemma 4.8 follows from (4.6).

\qed

Now we are ready to finish a proof of Proposition 4.5. Namely from (4.29) and
(4.31) it follows that $s(k,2,p)\in{\bf Q},~~(y_{r+1}+2y_r)\cdot s(k,2,p)
\in{\bf Z}$, and
$$s(0,2,p)={3p\over p+2}.$$
Finally, we observe that (4.27) follows from (4.29) and (4.31) if we replace
all modified Bernoulli polynomials by ordinary ones.

\qed
\medbreak

{\bf Proposition 4.9.}~~For all positive $p\in{\bf Q}$ the remainder term in
(4.27) lies in $6{\bf Z_+}$. More exactly, given a positive  $p\in{\bf Q}$
we define a set of integers $\{s_k\},~~k=0,1,2,\ldots $ such that
$$\left[{j+1\over p+2}\right] =k~~{\rm iff}~~s_k\le j<s_{k+1},~s_0:=0.
$$
Further, let us define a function
$$t(j):=t(j,p)=(2k+1)j+k-2\sum^k_{a=0}s_a ~~{\rm iff}~~s_k\le j<s_{k+1}.
$$
Then we have
$$s(j,2,p)={3p\over p+2}-{6j(j+2)\over p+2}+6t(j,p).
$$
\qed

It is clear that $t(j,p)\in{\bf Z}_+$.
\medbreak

{\bf Corollary 4.10.}~~Let us fix the positive integers $l=1,2,3,\ldots$ (the
fusion level), $p>q,~~j_1$ and $j_2$. Then
$$s\left( j_1,2,{ql\over p-q}-2\right) +s(r_0,2,l)-s\left( j_2,2,{pl
\over p-q}-2\right) =c-24\Delta +6{\bf Z},\eqno (4.32)
$$
where $r_0=l\cdot\left\{\displaystyle{{j_1-j_2\over l}}\right\}$ and
$$\eqalignno{
&c={3l\over l+2}\left( 1-{2(l+2)(p-q)^2\over l^2pq}\right),\cr
&~~&(4.33)\cr
&\Delta ={[p(j_1+1)-q(j_2+1)]^2-(p-q)^2\over 4lpq}+{r_0(l-r_0)\over
2l(l+2)}}
$$
are the central charge and conformal dimensions of RCFT, which corresponds
to the coset pair [GKO]
$$\matrix{
&&A_1&\oplus& A_1&\supset& A_1\cr
& & & \cr
{\rm levels}&&{\displaystyle ql\over\displaystyle p-q}-2&&l
&&{\displaystyle pl\over\displaystyle p-q}-2\cr}
$$
\vskip 1cm

{\bf\S 5.~~Proof~ of~Theorem~3.1.}

\bigbreak

We start with a generalization of identity (4.2).

\medbreak

{\bf Proposition~5.1.}~~For all real $\varphi,~\psi$ and $\theta$ we have

$$\eqalignno{
&L\Bigg ({\sin\theta\over\sin\varphi}\cdot{\sin (\theta +\psi )\over\sin
(\varphi+\psi )}\Bigg )=&(5.1)\cr
&=\pi^2\biggl\{ {\overline B}_2\biggl({\theta+\varphi+\psi\over\pi}\biggr)-
{\overline B}_2\biggl({2\theta + \psi +{\overline\epsilon}(\theta ,\theta +
\psi )\pi\over 2\pi}\biggr)-\cr
&-{\overline B}_2\biggl({2\varphi+\psi +{\overline\epsilon}(\varphi ,\varphi +
\psi )\pi\over 2\pi}\biggr)+{1\over 6}\biggr\} +\cr
&+L\biggl(-{\sin (\varphi -\theta )\over\sin\theta },\varphi\biggr)+
L\biggl(-{\sin (\varphi
-\theta )\over \sin (\theta +\psi )},\varphi +\psi \biggr)- \cr
&-L\biggl(-{\sin\varphi\over\sin (\theta  +\psi )},\varphi +\theta +\psi\biggr)
-L\biggl(-{\sin (\varphi +\psi )\over\sin\theta },\varphi +\theta
+\psi\biggr),  \cr }
$$
where ${\overline\epsilon}(\varphi ,\theta )=1-\epsilon (\varphi ,\theta )$
and $\epsilon (\varphi ,\theta )$ is defined by (4.31).

Proof.~~First of all we consider the case when $0<\varphi +\theta +\psi <\pi$
and $\varphi ,\theta ,\psi >0$. In this case ${\overline\epsilon}(\theta ,
\theta +\psi )={\overline\epsilon}(\varphi ,\varphi +\psi )=0$ and one can use
the identities (4.12)-(4.14) in order to show that a derivative of difference
between LHS and RHS of (5.1) with respect to $\varphi$ is equal to
$$2\pi\left\{ B_1\left({\theta +\varphi +\psi\over \pi}\right) -B_1\left(
{2\varphi +\psi\over 2\pi}\right)\right\} =2\theta +\psi .
$$
Integrating (see (4.10)) we find that the difference between LHS and RHS of
(5.1) is a function $c(\theta ,\psi )$ which does not depend on $\varphi$.
In order to find $c(\theta ,\psi )$ let us take $\varphi =\theta$ in (5.1).
After this substitution we obtain an equality
$$\eqalignno{
{\pi^2\over 6}=&{1\over 2}(2\theta +\psi )^2+c(\theta ,\psi )-\cr
&-L\left(-{\sin\theta\over\sin (\theta +\psi )},2\theta +\psi \right)-
L\left(-{\sin (\theta +\psi )\over\sin\theta},2\theta +\psi \right).}
$$
Comparing the last equality with (4.31) (in our case $\epsilon (\theta ,
\theta +\psi )=1$) we find $c(\theta ,\psi )=0$. In general case we use the
reduction rules (4.3), (4.4) and (4.6) and the following properties of
function $\epsilon (\varphi ,\theta )$:
$$\eqalignno{
&\epsilon (\theta ,\theta )=1, ~~~~~~~~\epsilon (\theta ,\varphi )=
\epsilon (\varphi ,\theta ),\cr
&\epsilon (\theta ,\pi +\theta )=0, ~~~\epsilon (\varphi ,-\theta )=
\epsilon (\varphi ,\pi +\theta ),\cr
&\epsilon (\theta ,\pi -\theta )=1, ~~~\epsilon (-\varphi ,-\theta )=
\epsilon (\varphi ,\theta ).}
$$
\qed
\medbreak

{\bf Corollary~5.2.}~~We have
$$L\left( {\sin(\varphi +\theta )\over\sin\theta},\varphi \right)+L\left( {\sin
(\varphi +\theta)\over\sin\varphi},\theta\right)=2\pi^2\biggl\{
{\overline B}_2\left(
{\varphi+\theta +{\overline\epsilon}(\varphi ,\theta )\pi\over 2\pi}\right)
 +{1\over 12}\biggr\}.\eqno (5.2)
$$
Proof.~~Take $\psi =-\theta -\varphi$ in (5.1).

\qed

Let us continue a proof of Theorem 3.1 and take a specialisation $\theta\to
 k\varphi ,\break\varphi\to (m+k)\varphi$ and $\psi\to (n-2k)\varphi$ in (5.1).
Then we obtain
$$\eqalignno{
&L\Bigg ({\sin k\varphi\over\sin (m+k)\varphi}\cdot{\sin (n-k)\varphi\over\sin
(m+n-k)\varphi}\Bigg )=\cr
&=\pi^2\biggl\{ {\overline B}_2\biggl({(m+n)\varphi\over\pi}\biggr)-
{\overline B}_2\biggl({n\varphi +{\overline\epsilon}(k\varphi ,(n-k)\varphi
 )\pi\over 2\pi}\biggr)-\cr
&-{\overline B}_2\biggl({(n+2m)\varphi+{\overline\epsilon}((m+k)\varphi ,
(n-k)\varphi )\pi \over 2\pi}\biggr)+
{1\over 6}\biggr\} +\cr
&+L\biggl(-{\sin m\varphi \over\sin k\varphi },(m+k)\varphi\biggr)+
L\biggl(-{\sin m\varphi
\over \sin (n-k)\varphi },(m+n-k)\varphi \biggr)- \cr
&-L\biggl(-{\sin (m+k)\varphi\over\sin (n-k)\varphi },(m+n)\varphi \biggr)
-L\biggl(-{\sin (m+n-k)\varphi \over\sin k\varphi },(m+n)\varphi \biggr). \cr }
$$
Consequently, after summation we obtain
$$\eqalignno{
&\sum_{k=1}^{n-1}\sum_{m=1}^rL\Bigg ({\sin k\varphi\over\sin (m+k)\varphi}
\cdot{\sin (n-k)\varphi\over\sin (m+n-k)\varphi}\Bigg )=
2\sum_{k=1}^{n-1}\sum_{m=1}^rL\biggl(-{\sin m\varphi \over\sin k\varphi },
(m+k)\varphi\biggr)- \cr
&-2\sum_{k=1}^{n-1}\sum_{m=1}^rL\biggl(-{\sin (m+n-k)\varphi
\over\sin k\varphi },(m+n)\varphi \biggr) +\pi^2\Sigma_3= \cr
&=2\sum_{k=1}^{n-1}\sum_{m=1}^{n-k}L\biggl(-{\sin m\varphi
\over\sin k\varphi },(m+k)\varphi\biggr)-
2\sum_{k=1}^{n-1}\sum_{m=1}^{n-k}L\biggl(-{\sin (r+m)\varphi
\over\sin k\varphi },(m+r+k)\varphi \biggr) +\cr
&~~~\cr
&+\pi^2\Sigma_3
=2\Sigma_1-2\Sigma_2+\pi^2\Sigma_3,~~~{\rm where}~~~ \cr
& ~~~ \cr
&\Sigma_3:=\sum_{k=1}^{n-1}\sum_{m=1}^r\biggl\{{\overline B}_2\biggl({(m+n)
\varphi\over\pi}\biggr)
-{\overline B}_2\biggl({n\varphi +{\overline\epsilon}(k\varphi ,(n-k)\varphi
 )\pi\over 2\pi}\biggr)- &(5.3)\cr
&-{\overline B}_2\biggl({(n+2m)\varphi+{\overline\epsilon}((m+k)\varphi ,
(n-k)\varphi )\pi \over 2\pi}\biggr)+
{1\over 6}\biggr\} .}
$$
At first, let us consider the sum
$$\eqalignno{
&2\Sigma_1:=2\sum_{k=1}^{n-1}\sum_{m=1}^{n-k}L\biggl(-{\sin m\varphi
\over\sin k\varphi },(m+k)\varphi\biggr)= 2\sum_{p=1}^{\left[{n\over 2}
\right]}L(-1,2p\varphi )+ &(5.4)\cr
&+2\sum_{p=3}^{n}\sum_{k=1}^{\left[{p-1\over 2}\right]}\biggl\{
L\biggl(-{\sin (p-k)\varphi
\over\sin k\varphi },p\varphi\biggr)+2L\biggl(-{\sin k\varphi
\over\sin (p-k)\varphi },p\varphi\biggr)\biggr\}= \cr
&=2\sum_{p=3}^n\sum_{k=1}^{\left[{p-1\over 2}\right] }2\pi^2{\overline B}_2
\left({p\varphi +{\overline\epsilon}((p-k)\varphi ,k\varphi )\pi\over 2\pi}
\right) +2\sum_{p=1}^{\left[{n\over 2}\right]}\pi^2{\overline B}_2\left(
{p\varphi\over\pi} +{1\over 2}\right). }
$$

Secondly, in order to compute the sum $2\Sigma_2$, let us remind that
$\displaystyle{\varphi ={(j+1)\pi\over n+r}}$. Hence $\displaystyle{\sin (m+r)
\varphi =\sin{(m+r)(j+1)\pi\over n+r}=(-1)^j\sin (n-m)\varphi}$ and
consequently\break (see (4.3) and (4.4))
$$\eqalignno{
&L\biggl( -{\sin (r+m)\varphi \over\sin k\varphi },(m+r+k)\varphi \biggr) =\cr
&=L\biggl( (-1)^{j+1}{\sin (n-m)\varphi \over\sin k\varphi },(j+1)\pi -
(n-m-k)\varphi \biggr) =\cr
&=L\biggl( -{\sin (m-n)\varphi \over\sin k\varphi },
(m+k-n)\varphi \biggr) .\cr}
$$
So we have
$$\eqalignno{
&2\Sigma_2:=2\sum_{k=1}^{n-1}\sum_{m=1}^{n-k}L\biggl(- {\sin (r+m)\varphi
\over\sin k\varphi },(m+r+k)\varphi \biggr) =& (5.5)\cr
&=\sum_{k=1}^{n-1}\sum_{m=1}^{n-k}L\biggl( -{\sin (m-n)\varphi
\over\sin k\varphi },(m+k-n)\varphi \biggr) =2\sum_{p=1}^{\left[ {n-1
\over 2}\right] }L(2\cos p\varphi ,p\varphi )+{(n-1)\pi^2\over 3}+\cr
&+2\sum_{p=3}^{n-1}\sum_{k=1}^{\left[ {p-1\over 2}\right] }\biggl\{ L\biggl(
{\sin p\varphi \over\sin k\varphi },(p-k)\varphi \biggr) +
L\biggl( {\sin p\varphi \over\sin (p-k)\varphi },k\varphi \biggr)\biggr\} =
{(n-1)\pi^2\over 3}+\cr
&+2\sum_{p=3}^{n-1}\sum_{k=1}^{\left[{p-1\over 2}\right] }2\pi^2\biggl\{
{\overline B}_2
\left( {p\varphi +{\overline\epsilon}(k\varphi ,(p-k)\varphi )\pi\over 2\pi}
\right) +{1\over 12}\biggr\} +2\sum_{p=1}^{\left[ {n\over 2}\right] }\pi^2
\biggl\{ {\overline B}_2\left( {p\varphi\over\pi} \right) +{1\over 12}
\biggr\} . \cr}
$$
Let us sum up our computations. First of all we proved that
$s(j,n,k)\in{\bf Q}$. Secondly, in order to compute the dilogarith sum
$s(j,n,k)$ modulo${\bf Z}$ we may replace all modified Bernoulli polinomials
appearing in (5.3)-(5.5) by ordinary ones. After some bulky calculations we
obtain (3.5), except positivity of a remainder term in (3.5). Finally, in
order to obtain the exact formulae (3.3) and (3.4) we are based on the
properties of Dedekind sums [Ra]. Details will appear elsewhere.

\qed

Now we propose a generalisation of (4.27). For this goal let us define the
following function
$$L_k(\theta ,\varphi ):=2L\left({\sin\theta\cdot\sin k\theta\over\sin\varphi
\cdot\sin (\varphi +(k-1)\theta )}\right)-\sum_{j=0}^{k-1}L\left(\left(
{\sin\theta\over\sin (\varphi +j\theta )}\right)^2\right).\eqno (5.6)
$$
\medbreak

{\bf Lemma 5.3.} We have
$$L_k(\theta ,\varphi )=2L\left( -{\sin (\varphi -\theta )\over\sin k\theta},~
\varphi +(k-1)\theta\right)-2L\left( -{\sin\varphi\over\sin k\theta},~\varphi
+k\theta\right) +\pi^2{\bf Q}.
$$
\qed

Now let $p\in{\bf Q}$ and consider a decomposition of $p/k$ into continued
fraction
$${p\over k}=b_r+{1\over\displaystyle b_{r-1}+
{\strut 1\over\displaystyle\cdots +{\strut 1\over\displaystyle b_1+
{\strut 1\over
\displaystyle b_0}}}},\eqno (5.7)
$$
where $b_i\in{\bf N},~~0\le i \le r-1$ and $b_r\in{\bf Z}$. Using the
decomposition (5.7) we define (compare with (4.25)):
$$\eqalignno{
&y_{-1}=0,~~y_0=1,~~y_1=b_0,\ldots ,y_{i+1}=y_{i-1}+b_iy_i,\cr
%~~0\le i\le r,
&m_0=0,~~m_1=b_0,~~m_{i+1}=|b_i|+m_i,\cr
%~~0\le i\le r.}
&r(j):=r_k(j)=i,~~~{\rm if}~~km_i\le j<km_{i+1}+\delta_{i,r},\cr
%~~0\le i\le r,\cr
&n_j:=n_k(j)=ky_{i-1}+(j-km_i)y_i,~~{\rm if}~~km_i\le j<km_{i+1}+\delta_{i,r},
%~~0\le i\le r.
}
$$
where $0\le i\le r$.

Finally, we consider the following dilogarithm sum
$$\sum_{j=1}^{km_{r+1}}(-1)^{r(j)}L_k\bigg( y_{r(j)}\theta ,~(n_j+y_{r(j)})
\theta\bigg) =(-1)^r{\pi^2\over 6}s(l,k+1,p),\eqno (5.8)
$$
where ${\displaystyle \theta ={(l+1)\pi\over ky_{r+1}+(k+1)y_r}}$.
\medbreak

{\bf Proposition 5.4.} We have
$$\eqalignno{
&(i)~~~s(0,k+1,p):=c_k={3(p+1-k)\over p+k+1}, ~~k\ge 1,&(5.8)\cr
&(ii)~~s(l,k+1,p)=c_k-{6k~l(l+2)\over p+k+1}+6{\bf Z}, &(5.9)}
$$
$(iii)$ if $k=1$ or $2$, then the remainder term in (5.9) lies in $6{\bf Z}_+$.

%-------------------------------------------------------------------
%\end k1
%
%-------------------------------------------------------------------
%
%\input rf
%
%-------------------------------------------------------------------
\vskip 1cm

{\bf Riferences.}

\medbreak

\item{[AL]} D.Acreman, J.H.Loxton, Aeq. Math., 30, (1986), 106.
\item{[Bi]} A.Bilal, Nucl. Phys., B330, (1990), 399.
\item{[Bl]} S.Bloch, Application of the dilogarithm function in algebraic
$K$-theory and algebraic geometry, Proc. of the International Symp. on Alg.
Geom.:(Kyoto Univ., Kyoto, 1977) 103-114, Kinokuniya Book Store, Tokyo, 1978.
\item{[BPZ]} A.Belavin, A.Polyakov, A.Zamolodchikov, J. Stat. Phys., 34,
(1984), 763; Nucl. Phys., B241, (1984), 333.
\item{[BR]} V.V.Bazhanov, N.Yu.Reshetikhin, J. Phys. A, 23, (1990), 1477.
%Prog.Theor.Phys., 102, (1990), suppl.
\item{[CR]} P.Christe, F.Ravanini, $G_N\otimes G_L/G_{N+L}$ conformal field
theories and their modular invariant partition functions, Int. J. Mod. Phys.,
A4, (1989), 897.
\item{[DR]} P.Dorey, F.Ravanini, Staircase model from affine Toda field
theory, prepr.\break SPhT/92-065.
\item{[FF]} B.L.Feigin, D.B.Fuks, Springer Lecture Notes in Mathematics,
1060, (1984), 230.
\item{[FL]} V.A.Fateev, S.Lukyanov, Int. J. Mod. Phys., A3, (1988), 507.
\item{[FQS]} D.Friedan, Z.Qiu, S.Shenker, Phys. Rev. Lett., 52, (1984), 1575.
\item{[FS]} E.Frenkel, A.Szenes, Dilogarithm identities, $q$-difference
equations and the Virasoro algebra, prepr. hep-th/9212094.
\item{[GKO]} P.Goddard, A.Kent, D.Olive, Phys. Lett., B152, (1985), 105.
\item{[GM]} I.M.Gelfand, R.D.MacPherson, Adv. in Math., 44, (1982), 279.
\item{[Ka]} V.G.Kac, Infinite dimensional Lie algebras, Cambridge Univ. Press.
1990.
\item{[KBP]} A.Klumper, M.T.Batchelor, P.A.Pearce, J. Phys. A: Math. Gen., 24,
(1991), 311.
\item{[Ki1]} A.N.Kirillov, Zap. Nauch. Semin. LOMI, 164, (1987), 121.
\item{[Ki2]} A.N.Kirillov, Dilogarithm identities and spectra in conformal
field theory, Talk given at the Isaac Newton Institute, Cambridge,
October 1992, prepr. hep-th/9211137.
\item{[KKMM]} R.Kedem, T.R.Klassen, B.M.McCoy, E.Melzer, Fermionic
quasi-particle representations for characters of $(G^{(1)})_1\times (G^{(1)})
_1/(G^{(1)})_2$, prepr. ITP-SB-92-64.
\item{[Kl]} A.Klumper, Ann. Physik, 1, (1992), 540.
\item{[KM]} R.Kedem, B.M.McCoy, Construction of modular branching functions
from Bethe's equations in the 3-state Potts chain, prepr. ITP-SB-92-56.
\item{[K-M]} T.Klassen, E.Melzer, Nucl. Phys., B370, (1992), 511.
\item{[KN1]} A.Kuniba, T.Nakanishi, Level-rank duality in fusion RSOS models,
prepr., 1989.
\item{[KN2]} A.Kuniba, T.Nakanishi, Spectra in conformal field theories from
Rogers dilogarithm, prepr. SMS-042-92, 1992.
\item{[KP]} A.Klumper, P.Pearce, Physica, A183, (1992), 304.
\item{[KR1]} A.N.Kirillov, N.Yu.Reshetikhin, Zap. Nauch. Semin. LOMI, 160,
(1987), 211.
\item{[KR2]} A.N.Kirillov, N.Yu.Reshetikhin, J. Phys. A. Math. Gen., 20,
(1987), 1565.
\item{[Ku]} A.Kuniba, Thermodynamics of the $U_q(X_r^{(1)})$ Bethe ansatz
system with $q$ a root of Unity, ANU, prepr. SMS-098-91.
\item{[KW]} V.Kac, M.Wakimoto, Adv. Math., 70, (1988), 156.
\item{[Le1]} L.Lewin, J. Austral. Math. Soc., A33, (1982), 302.
\item{[Le2]} L.Lewin, Polylogarithms and associated functions, (North-Holland,
1981).
\item{[Lo]} J.H.Loxton, Acta Arithmetica, 43, (1984), 155.
\item{[Mi]} J.Milnor, Hyperbolic geometry : the first 150 years, Bull A.M.S.,
6, (1982), 9-24.
\item{[MSW]} P.Mathieu, D.Senechal, M.Walton, Field identification in
nonunitary diagonal cosets, prepr. LETH-PHY-9/91.
\item{[Na]} W.Nahm, Dilogarithm and W-algebras, Talk given at the Isaac
Newton Inst., Cambridge, Sept. 1992.
\item{[NRT]} W.Nahm, A.Recknagel, M.Terhoeven, Dilogarithm identities in
conformal field theory, BONN-prepr., hep-th/9211034.
\item{[NZ]} W.D.Neumann, D.Zagier, Topology, 24, (1985), 307.
\item{[Ra]} H.Rademacher, Lectures on Analytic Number Theory (Notes,
Tata Inst. of Fundamental Research, Bombay, 1954-1955).
\item{[Ro]} L.J.Rogers, Proc. London Math. Soc., 4, (1907), 169.
\item{[RS] } B.Richmond, G.Szekeres, J. Austral. Math. Soc., A31, (1981), 362.
\item{[SA]} H.Saleur, D.Altshuler, Nucl. Phys., B354, (1991), 579.
\item{[Te]} M.Terhoeven, Lift of dilogarithm to partition identities, prepr.
Bonn-HE-92-36.
\item{[Th]} W.P.Thurston, The Geometry and Topology of 3-manifolds (Princeton
University Press, 1983).
\item{[TS]} M.Takahashi, M.Suzuki, Prog. Theor. Phys., 48, (1972), 2187.
\item{[Wa]} G.N.Watson, Quart. J. Math., Oxford, ser. 8, (1937), 39.
\item{[Y]} T.Yoshida, Invent. Math., 81, (1985), 473.
\item{[Z]} Al.B.Zamolodchikov, Nucl. Phys., B342, (1990), 695.
\item{[Za]} D.Zagier, The remarkable dilogarithm (Number theory and related
topics. Papers presented at the Ramanujan colloquium, Bombay, 1988, TIFR).

\end